\documentclass{llncs}
\usepackage{graphicx}
\usepackage{url}
\begin{document}

\mainmatter              % start of the contributions
%
%\title{On Choice Overload in Recommendations in Digital Libraries}
%\title{Choice Overload in Digital Libraries:\\Are Researchers Overstrained When Receiving More Than One Recommendation at a Time?}
\title{Exploring Choice Overload in Related-Article Recommendations in Digital Libraries}
\author{Felix Beierle\inst{1} \and Akiko Aizawa\inst{2} \and Joeran Beel\inst{2,} \inst{3}}
\authorrunning{Felix Beierle, Akiko Aizawa, Joeran Beel} % abbreviated author list (for running head)
%
%%%% list of authors for the TOC (use if author list has to be modified)
\tocauthor{Felix Beierle, Akiko Aizawa, Joeran Beel}
\institute{
	Service-centric Networking\\
	Technische Universit\"at Berlin / Telekom Innovation Laboratories\\
	Berlin, Germany\\
	\email{beierle@tu-berlin.de}\\
\and
	National Institute of Informatics (NII)\\
    Digital Content and Media Sciences Research Division\\
    Tokyo, Japan\\
	\email{\{aizawa,beel\}@nii.ac.jp}    
\and
	Trinity College Dublin\\
	School of Computer Science and Statistics\\
	Intelligent Systems Discipline, Knowledge and Data Engineering Group\\
	ADAPT Centre\\
	Dublin, Ireland\\
	\email{joeran.beel@adaptcentre.ie}
}

\maketitle              % typeset the title of the contribution

\begin{abstract}
%The abstract should summarize the contents of the paper
%using at least 70 and at most 150 words. It will be set in 9-point
%font size and be inset 1.0 cm from the right and left margins.
%There will be two blank lines before and after the Abstract. \dots
%OLD
%We tackle the question of how many related-articles to display when a user is viewing an item from a digital library. After investigating how existing digital libraries are handling it, we provide the results of an empirical evaluation, showing lower click-through rates for higher numbers of recommendations as well as twice as many clicked recommendations when displaying ten instead of one related-articles.
%OLD end
%
%We investigate the problem of choice overload when displaying
%recommendations in digital libraries.
%After analyzing how many items existing libraries display,
%we provide the results of an empirical evaluation, showing lower click-through rates
%for higher numbers of recommendations as well as twice as many clicked recommendations
%when displaying ten instead of one related-articles.
%Our results show that users might quickly feel overloaded by choice.
%Furthermore, we show that the quality of the recommendations has a significant impact
%on the click-through rates.
%
We investigate the problem of choice overload -- the difficulty of making a decision when faced with many options -- when displaying related-article recommendations in digital libraries. So far, research regarding to how many items should be displayed has mostly been done in the fields of media recommendations and search engines. We analyze the number of recommendations in current digital libraries. When browsing fullscreen with a laptop or desktop PC, all display a fixed number of recommendations. 72\% display three, four, or five recommendations, none display more than ten. We provide results from an empirical evaluation conducted with \emph{GESIS}' digital library \emph{Sowiport}, with recommendations delivered by recommendations-as-a-service provider \emph{Mr.\ DLib}. We use click-through rate as a measure of recommendation effectiveness based on 3.4 million delivered recommendations. Our results show lower click-through rates for higher numbers of recommendations and twice as many clicked recommendations when displaying ten instead of one related-articles. Our results indicate that users might quickly feel overloaded by choice.
%, or the relevance of the recommendations was so low that many users did not click further recommendations after clicking the first one.
%
%
%number of recommendations in current digital libraries
%all libraries display a fixed number of recommendations
%none displays more than ten
%and 74\% display three or five recommendations.
%
%
% * <j@beel.org> 2017-01-27T16:58:20.968Z:
% 
% you should efinatily add some information about the other analysis, i.e. the number of recommendations in current digital libraries. such information is very likely to be cited.  important is that all libraries display a fixed number of recommendations, none displys more than 10 and xxx% display so-and-so-many.
% 
% ^ <beierle@tu-berlin.de> 2017-01-31T15:35:27.193Z.
% * <j@beel.org> 2017-01-27T16:35:25.916Z:
% 
% > might quickly feel overloaded by choice.
% we probably should be self critical because the results  are not very good. so, maybe you add "..., or the relevance of the recommendations was so low that many users did not click further recommendations after clicking the first one"
% 
% ^ <beierle@tu-berlin.de> 2017-01-31T15:35:39.393Z.
%
%\keywords{computational geometry, graph theory, Hamilton cycles}
\keywords{recommendation, recommender system, recommendations as a service, digital library, choice overload}
\end{abstract}

\section{Introduction}

%NEW

More and more information is available online for academic researchers in digital libraries
\cite{Noorden2014}.
One way to deal with the flood of information is to utilize recommender systems that
filter information and recommend articles related to those ones a user liked previously or is currently reading. A major challenge in recommending a list of related articles is to decide how many related articles to recommend before a user becomes dissatisfied with the recommender system due to choice overload.

\begin{figure}
	\centering
	\includegraphics[width=0.86\columnwidth]{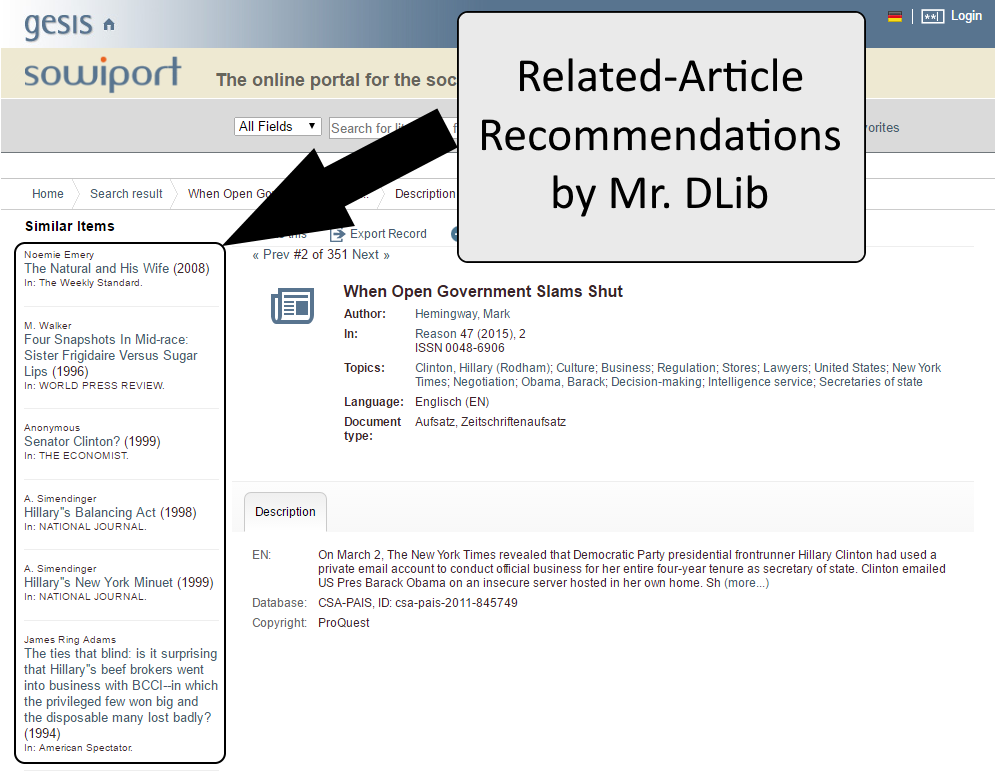}
	\caption{Screenshot of the Sowiport Digital Library showing related items on the left hand side.}
	\label{fig:screenshot}
\end{figure}

Developing Mr.\ DLib (Machine-readable Digital Library)\footnote{\url{http://mr-dlib.org}}
\cite{Beel2011b,Beel2017g}, a recommendations-as-a-service (RaaS)
provider,
we currently deliver recommendations to the digital library
Sowiport\footnote{\url{http://sowiport.gesis.org}}\cite{hienert_digital_2015}.
Soon, we will also deliver recommendations to
JabRef\footnote{\url{http://www.jabref.org}}\cite{Feyer2017}.
% * <j@beel.org> 2017-01-27T16:38:17.892Z:
% 
% this sentence is not understandable. too long, too many ( ), ...
% 
% ^ <beierle@tu-berlin.de> 2017-01-31T14:46:50.978Z.
%
Developing such a recommender system for digital libraries,
currently, there are no information to be found about how many recommendations to
deliver and display.
Figure \ref{fig:screenshot} shows an example of using Mr.\ DLib in Sowiport.
While a recommender system can
filter for the most relevant content for the user,
%give users the
% * <j@beel.org> 2017-01-27T16:39:38.663Z:
% 
% > While a recommender system can give users the
% > possibility to just display relevant content,
% 
% not clear. that sounds as if there was the option to display non-relevant content. 
% 
% ^ <beierle@tu-berlin.de> 2017-01-31T14:51:46.950Z.
%possibility to just display relevant content,
the displayed recommended items can still be overwhelming.
Schwartz describes the issue as the
"tyranny of choice" \cite{schwartz_tyranny_2004}:
Confronted with too many options, participants in studies tend to not
decide for any option.
In this paper, we investigate the influence of the size of the recommendation set
on choice overload in digital libraries.
In order to do so, we:
\begin{enumerate}
	\item Examine how many items other recommender systems in digital libraries recommend
	\item Conduct an empirical evaluation to see how different numbers of recommendations
	affect the clicks on related-article recommendations\footnote{
		All data  relating  to  this paper  is available  on 
		\url{http://datasets.mr-dlib.org},
% * <beierle@tu-berlin.de> 2017-03-21T14:14:40.033Z:
% 
% > Harvard's Dataverse via 
% > 		\url{http://datasets.mr-dlib.org}
% 
% Should the data be referenced with a DOI like in the ISI paper?
% 
% ^.
		including  a table of  the  delivered and  clicked  recommendations,
		the information about the investigated digital libraries,
		and the figures presented in this paper.}
\end{enumerate}

%we analyze CTR for a rec sys in order to:
%- see if results from entertainment is also applicable for the field of DLs
%-- see if choice overload can be seen
%to set up our experiment, we first analyze what other DLs are doing

%\begin{enumerate}
%	\item Look how other recommender systems are doing it
%	\item Conduct an empirical evaluation to see how different numbers of recommendations affect the effectiveness of related-article recommendations
%\end{enumerate}

\section{Related Work}

%general choice overload paper:
%Arunachalam: An Empirical Investigation into the Excessive-Choice Effect
%Fasolo: Escaping the tyranny of choice
There have been several studies investigating choice overload
with respect to consumer goods (for example
\cite{arunachalam_empirical_2009,fasolo_escaping_2007}).
%
% Bollen et al. 2010
Based on the MovieLens dataset, Bollen et al.\ investigated
the relationship between
\emph{item set variety},
\emph{item set attractiveness},
\emph{choice difficulty}, and
\emph{choice satisfaction}
\cite{bollen_understanding_2010}.
They suggest to diversify
the recommendation set by
including some lower quality recommended items
in order to increase perceived recommendation variety and
choice satisfaction.
%
% Willemsen et al. 2016
In another study, Willemsen et al.\ further analyzed
the relationship of diversification,
choice difficulty, and satisfaction \cite{willemsen_understanding_2016}.
Here, the authors also used the MovieLens dataset.
%

%SERP
Other related studies looked into the number of search results to be displayed.
%The screen size and choice overload.
%
%Kelly and Azzopardi studied the number of search results to be shown
%\cite{Kelly:2015:MRP:2766462.2767732}.
%In their study, they used three, six, and ten search results. One of their main
%findings is that subjects of the study who were shown ten search results per
%page viewed and saved significantly more documents, while more time is spent
%on earlier search results, if the number of results per page is less.
%
%Oulasvirta et al.\ describe choice overload as the "paradox of choice" when using search engines
%\cite{Oulasvirta:2009:MLP:1571941.1572030}.
%The authors conducted a user study with 24 participants and looked
%into the user's satisfaction with the results when displaying six or 24 results.
%For future work, they suggest also looking into
%objective behavior measurements like click-through rates.
%
%
Jones et al.\ conclude that the screen size is a determining factor
with respect to how many search result items users interact with
\cite{jones_improving_1999},
which is confirmed in a newer study by Kim et al.\ \cite{kim_eye-tracking_2015}.
Linden reports that although Google users claimed to
want more search results, traffic dropped with the
display of an increased number of search results
\cite{linden_marissa_2006}. Google suspected the extra loading time
to be play a role in this.
Azzopardi and Zuccon
developed a cost model for browsing search results, taking
into account screen size and search results page size \cite{azzopardi_two_2016}.
They concluded that displaying 10 results is close to the minimum cost.
Kelly and Azzopardi studied the effects of displaying different sizes of search
result pages \cite{Kelly:2015:MRP:2766462.2767732}.
In their study, they used three, six, and ten search results. One of their main
findings is that subjects of the study who were shown ten search results per
page viewed and saved significantly more documents, while more time is spent
on earlier search results, if the number of results per page is less.
While Chiravirakul and Payne's study suggests that
choice dissatisfaction happens when there is a lack of time for choosing links
\cite{chiravirakul_choice_2014},
the study by Oulasvirta et al.\ suggests that it is the search result page size
that is causing choice overload or the "paradox of choice."
%Oulasvirta et al.\ describe the choice overload as the "paradox of choice."
Oulasvirta et al.\ conducted a user study with 24 participants and looked
into the user's satisfaction with the results when displaying six or 24 results.
For future work, they suggest also looking into
objective behavior measurements like click-through rates.

%OUR work:
%-- domain is different
%-- mostly user studies
%-- we investigate if this is reflected in click rates
%-- tested in real recommender system actually deployed
%-- to set up our experiment, we first analyze what other DLs are doing

In our work, we focus on a different domain.
We will investigate the problem of choice overload
in digital libraries -- in contrast to movie recommendations and search results.
The question of how many recommendations to display and choice overload
has not been studied in the domain of digital libraries,
to the best of our knowledge.
In a recent literature survey on more than 200 articles about research paper recommender systems,
no one discussed or researched this topic \cite{Beel2014a}.
Furthermore, instead of user interviews or small scale studies, we
investigate the real clicks logged in an actively used system.
We will explore, to what extend click-through rates (CTR)
reflect the findings of the cited studies.

\section{Methodology}
\label{sec:method}
In order to investigate choice overload in digital libraries,
we first examine how many items other recommender systems in digital libraries display.
We investigated 63 digital libraries\footnote{Most of them listed on
\url{https://en.wikipedia.org/wiki/List_of_digital_library_projects}}
and reference managers with search interfaces\footnote{See \url{http://datasets.mr-dlib.org} for detailed results.}.
We considered recommendations that are being displayed when an item from the search
results is selected. In a few cases, the number of displayed recommendations of
related items was dependent on the size of the browser window. For the numbers given
in the following, we assumed a full screen browser window on a laptop computer
(13" display with 1280x800 resolution).

In a second step, we conducted an experimental evaluation to investigate how different
numbers of recommendations affect the click rates on related-article recommendations.
Click-through rates are a good way to study the users' actual behavior when displaying
recommended items in real situations.
%For this study, we analyze collected data from more than 3.4 million recommendations.
We analyze data from 3.4 million recommendations.
% * <beierle@tu-berlin.de> 2017-03-21T14:06:52.465Z:
% 
% > We analyze data from 3.4 million recommendations.
% I cannot see from the data I have, from what timeframe the data is. Can we add the daily average of delivered recommendations here for a frame of reference? What is the daily number?
% 
% ^.
The data was obtained from users of the academic
% * <j@beel.org> 2017-01-27T16:44:58.734Z:
% 
% >  For this study, we analyze collected data from
% > more than 3.4 million recommendations.
% 
% You often have quite a bit of redundancy in your text. For instance, instead of the above sentence you could just write "We analyze data from 3.4 million recommendations". In my opinion this short sentences contains the same amount of information but is much shorter. There are several other sentences like this in the text.
% 
% ^ <beierle@tu-berlin.de> 2017-01-31T14:54:00.726Z.
search engine Sowiport\footnote{Some explanations about Sowiport and Mr.\ DLib are from \cite{Beel2017d}.},
which is run by
\emph{GESIS - Leibniz-Institute for the Social Sciences}\footnote{\url{http://www.gesis.org}},
which is the largest infrastructure institution for the Social Sciences in Germany.
Sowiport contains about 9.6 million literature references and 50,000 research projects
from 18 different databases, mostly relating to the social and political sciences.
Literature references usually cover keywords, classifications, author(s), and journal
or conference information, and if available: citations, references, and links to full texts.

\begin{figure}[ht]
	\centering
	\includegraphics[width=0.975\columnwidth]{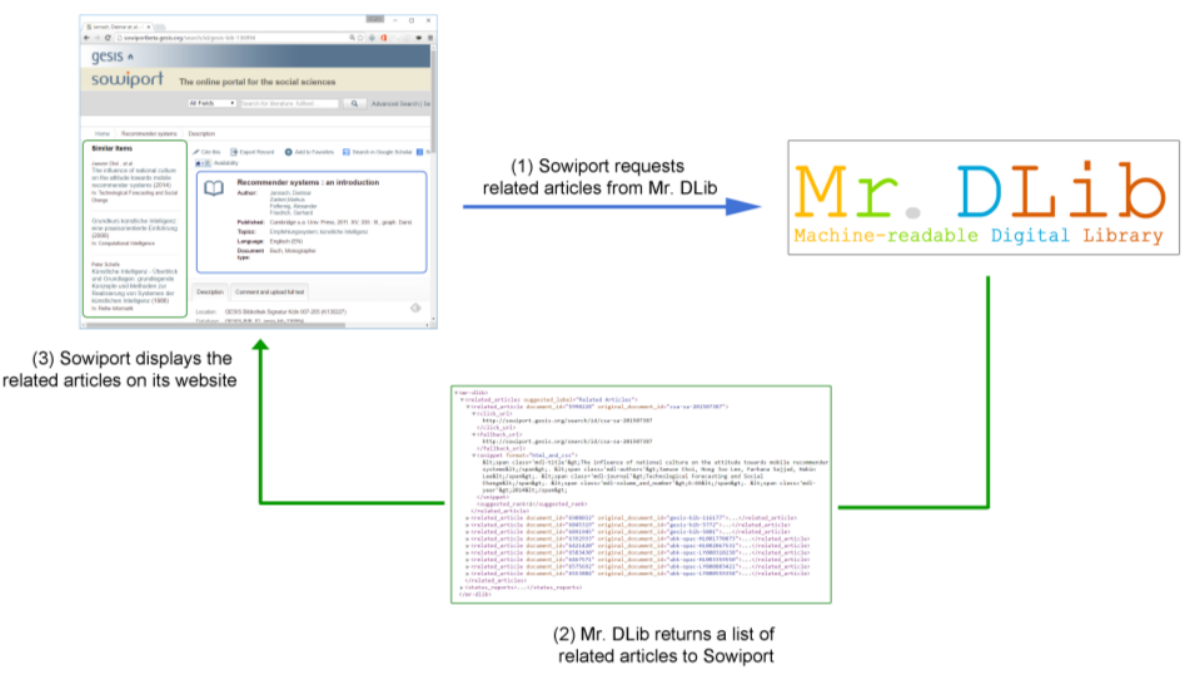}
	\caption{The recommendation process of Sowiport and Mr.\ DLib.}
	\label{fig:mrdlib-sowiport-process}
\end{figure}

Sowiport co-operates with Mr.\ DLib, an open Web Service to provide scholarly literature-recommendations-as-a-service (Figure \ref{fig:mrdlib-sowiport-process}).
This means that all computations relating to the recommendations run on Mr.\ DLib’s servers, while the presentation takes place on Sowiport’s website.
Our recommender system shows related-article recommendations on each article’s detail page in Sowiport (see Figure \ref{fig:screenshot}). Whenever such a detail page is requested by a user, the recommender system randomly chooses one of four recommendation approaches to generate recommendations: 1. stereotype recommendations, 2. most popular recommendations, 3. content-based filtering (CBF), and 4. random recommendations. 
% * <beierle@tu-berlin.de> 2017-03-21T14:05:46.322Z:
% 
% > . 
% Add reference here to a paper that deal with the four recommenders. Which paper is it?
% 
% ^.
We measured the effectiveness of the recommendation approaches with click-through rate (CTR). CTR describes the ratio of delivered to clicked recommendations. For instance, when 1,000 recommendations were delivered, and 8.4 of these recommendations were clicked, the average CTR would be 8.4/1,000=0.84\%. The assumption is that the higher the CTR, the more effective is the recommendation approach. There is some discussion to what extend CTR is appropriate for measuring recommendation effectiveness, but overall it has been demonstrated to be a meaningful and well-suited metric \cite{joachims_accurately_2005,beel_comparison_2015,schwarzer_evaluating_2016}.
% * <beierle@tu-berlin.de> 2017-03-21T14:10:09.953Z:
% 
% > There is some discussion to what extend CTR is appropriate for measuring recommendation effectiveness, but overall it has been demonstrated to be a meaningful and well-suited metric
% 
% From Review 2: "The goal should be to increase the usage and usefulness of the library. It is not clear at all why would "click-through rate" be a relevant model in this context. This measure would be important if one were to decide for example how many ads to pay for in order to get ROI."
% 
% We cite three papers indicating the usefulness of CTR for recommendation effectiveness. Is that enough or do we need to expand on this? If so, how?
% 
% ^.
%
For our evaluation, we randomly displayed one to fifteen recommendations.

% Expectations

\begin{figure}[ht]
	\centering
	\includegraphics[width=0.8\columnwidth]{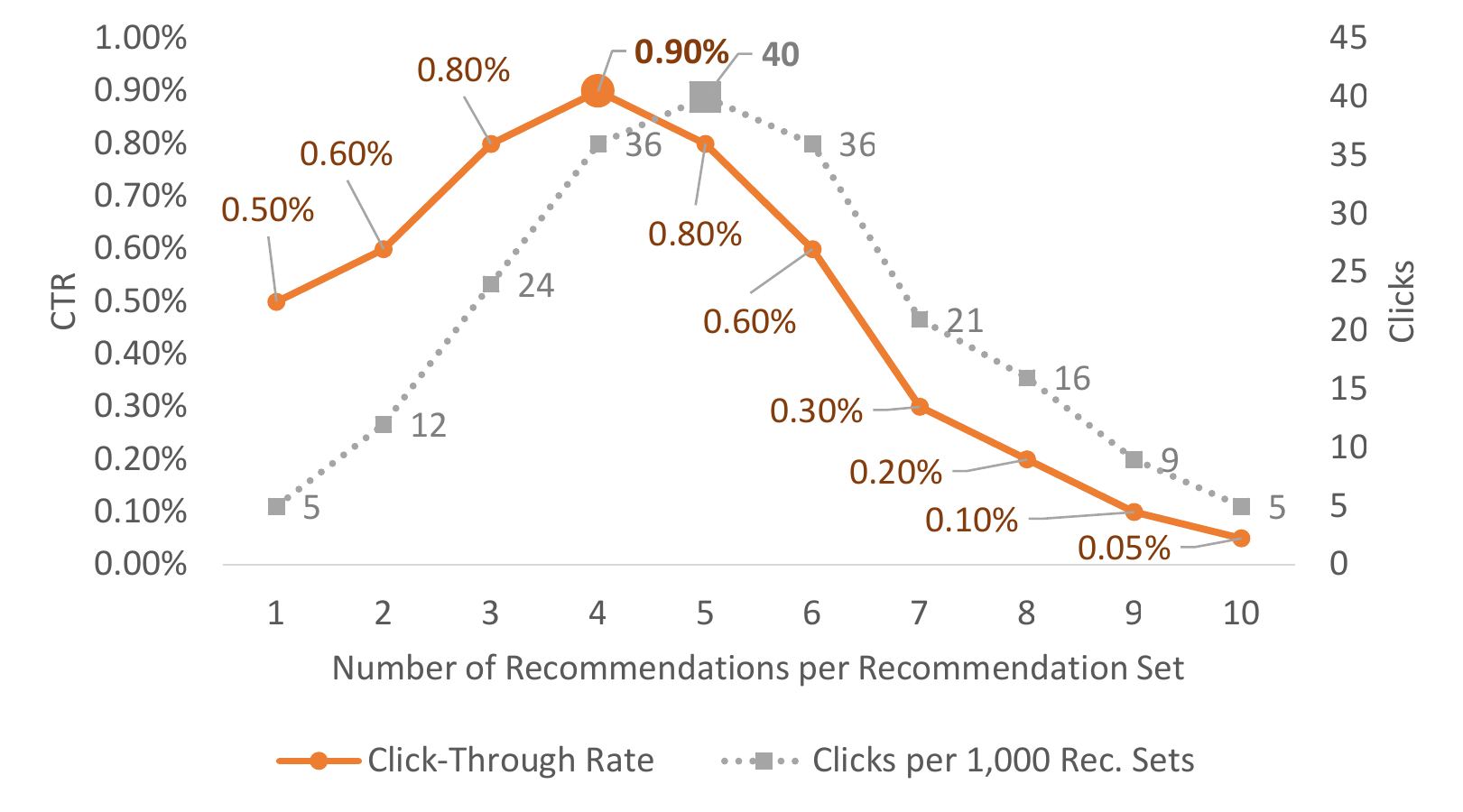}
	\caption{Expected click-through rate and clicks by displayed number of recommendations.}
	\label{fig:expectationa}
\end{figure}

Our expectation is that
at first, by increasing the number of displayed recommendations,
there will be an increase in the CTR.
By displaying more and more recommendations,
we expect to reach a maximum in CTR at some point.
After that, when displaying more recommendations,
we expect the CTR to drop, indicating choice overload.
Similarly, for the clicks, we would expect to find a maximum
at a certain number of displayed recommendations.
Plotting the CTR and the average clicks we show this expectation in Figure \ref{fig:expectationa}.
Here, the CTR, the orange line, increases with the number of displayed recommendations,
reaches a maximum at four, and decreases afterwards, indicating choice overload.
The clicks, the gray line, reach a maximum at five displayed recommendations.
In that case, we would decide for four (maximum CTR) or five (maximum clicks) recommendations.

% * <j@beel.org> 2017-01-27T16:49:13.380Z:
% 
% > Our expectation is t
% 
% i dont think reviewers will understand our expectations. the following paragraphs need a clearer structure, and more detailed explanations of the charts.
% 
% ^ <beierle@tu-berlin.de> 2017-02-07T16:25:12.382Z.
%Our expectation is that in general, displaying twice as many recommendations
%will also double the clicks.
%After a certain number of displayed related-article recommendations,
%the ratio will probably drop,
%e.g., displaying twice as many recommendations but only getting 1.5 times as many clicks.
%This would indicate choice overload and indicate to stop delivering and displaying more
%recommendations.
%Figure \ref{fig:expectationa} visualizes this expectation.
%Here, the CTR would increase with the number of displayed recommendations,
%reach a maximum at four, and decrease afterwards, indicating choice overload.
%In that case, we would decide for four (maximum CTR) or five (maximum clicks) recommendations.

\begin{figure}[ht!]
	\centering
	\includegraphics[width=0.8\columnwidth]{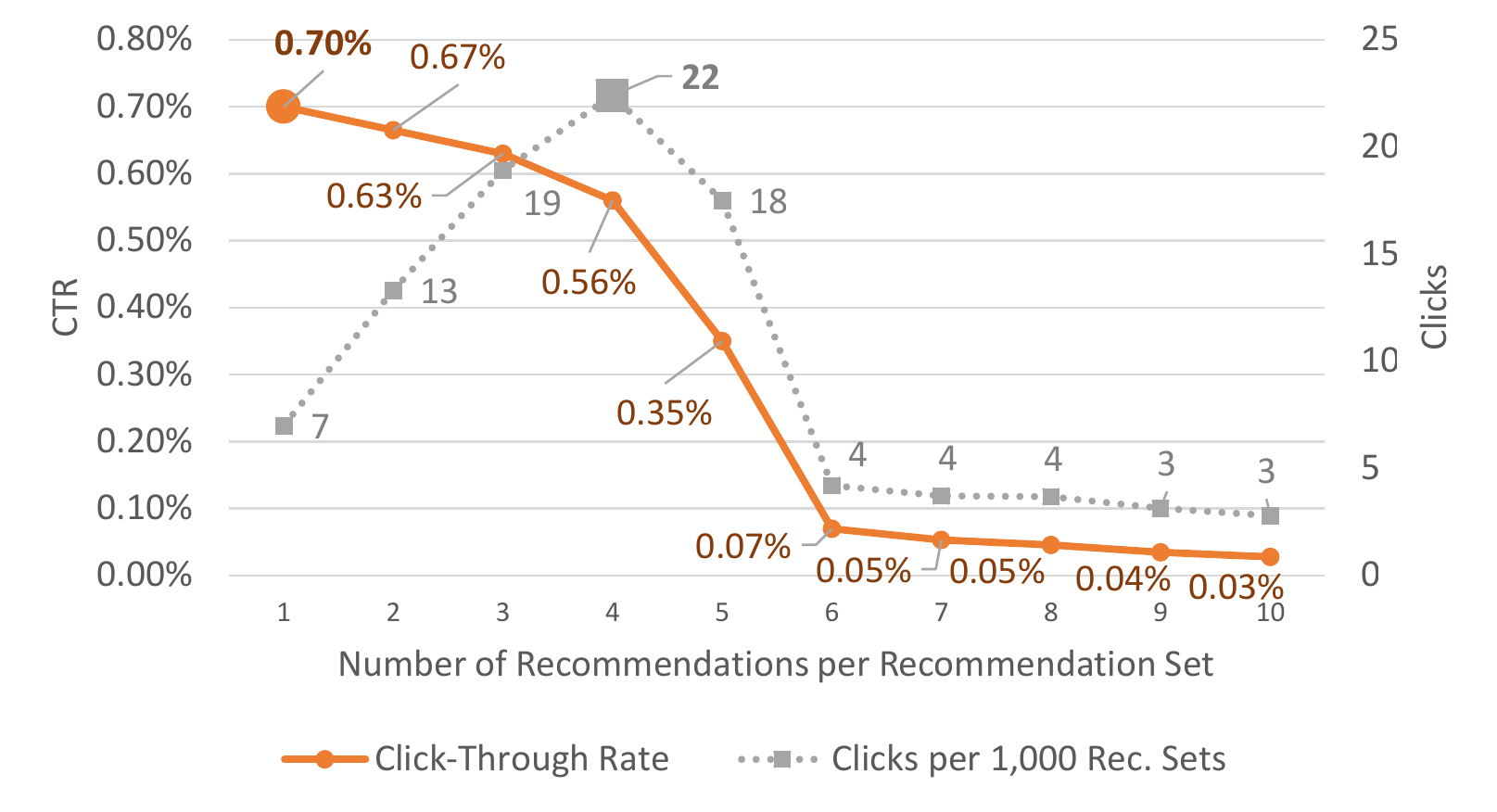}
	\caption{Alternative expected click-through rate and clicks by displayed number of recommendations.}
	\label{fig:expectationb}
\end{figure}

Alternatively, we would have expected results as in Figure \ref{fig:expectationb}.
Here, the CTR declines with increasing number of displayed recommendations.
The absolute clicks have a maximum at four.
The question then would be: What is better -- a higher CTR or the maximum
of absolute clicks?
Depending on the answer, we would choose a
recommendation set size of between one and four.
As we will show in the following section, the results are quite different from our expectations.


%\newpage
\section{Results}

\subsection{Number of Recommendations in Existing Digital Libraries}
\begin{figure}
	\centering
	\includegraphics[width=0.8\columnwidth]{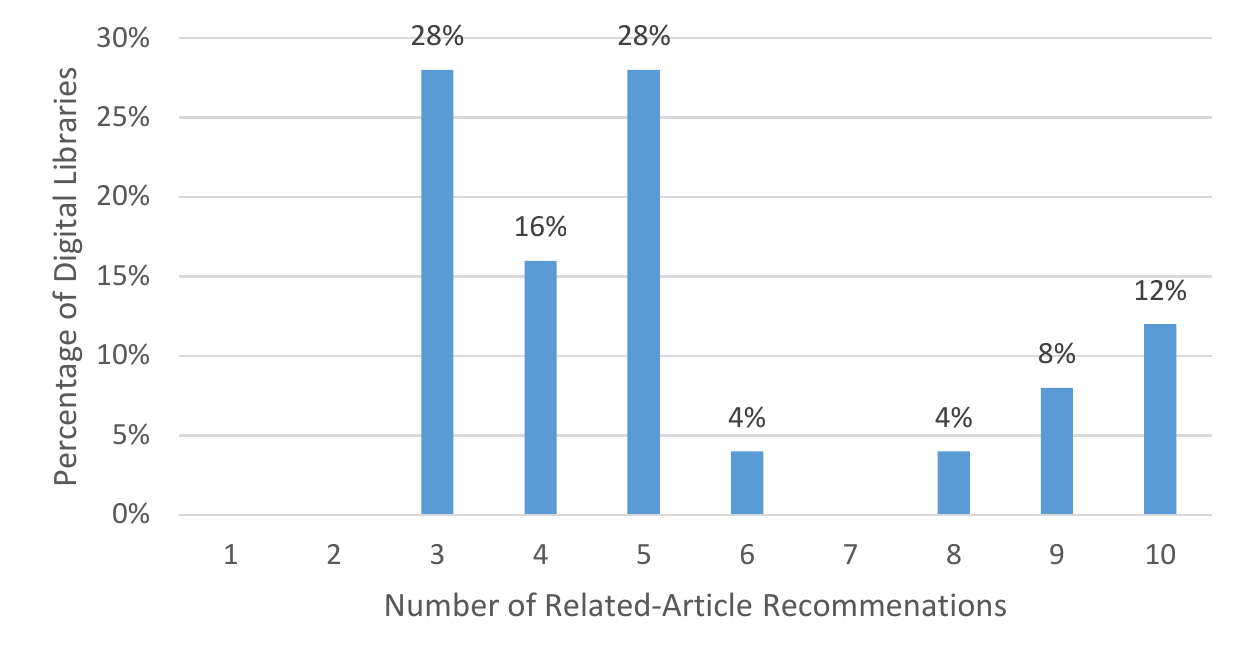}
	\caption{Number of displayed recommendations in current digital libraries.}
	\label{fig:stats}
\end{figure}

% * <j@beel.org> 2017-01-27T16:50:40.424Z:
% 
% > Existing Approaches}
% Better: "Number of Recommendations in Existing Digital Libraries"
% 
% ^ <beierle@tu-berlin.de> 2017-01-31T14:55:37.045Z.
19 (30\%) of the 63 digital libraries displayed recommendations for
% * <j@beel.org> 2017-01-27T16:51:29.263Z:
% 
% > investigated
% another example of  too much text. you can just delete "investigated" and everybody will still well understand the sentence. 
% 
% ^ <beierle@tu-berlin.de> 2017-01-31T14:55:35.463Z.
related items. Figure \ref{fig:stats} shows the distribution of recommendations for
those 19 libraries. Most libraries (72\% of those that display recommendations) display three, four, or five recommendations, none is
displaying more than 10 or less than three.
%
%http://www.charlesrcook.com/archive/2012/08/31/how-to-export-excel-plots-to-a-vector-image-eps.aspx
%
It is also notable that they all always show a fixed number of related-articles,
one could also imagine displaying varying numbers.
%based on a relevance score of similar items in the database.
%
Looking for related items in the database, there can be varying scores
of relevance.
One way of displaying a varying amount of related-article recommendations
would be to take into account such relevance scores,
e.g., giving 10 recommendations if there are 10 highly relevant related-articles,
and giving only two recommendations, if there are only two related-articles
that have a relevance score above a certain threshold.

We do not know how the numbers of related articles in the reviewed recommender systems
were chosen by the operators.
We assume that they either just arbitrarily chose the numbers, or did some
experiments but did not publish the results.
In the following, by measuring CTRs, we want to investigate and measure the effect the number of displayed
recommendations has and if the CTR indicates choice overload for certain numbers of
displayed recommendations.

\subsection{Experiment with Varying Number of Recommendations}
% * <j@beel.org> 2017-01-27T16:53:02.625Z:
% 
% > Data Analysis}
% Better: "Experimen with a Varying Number of Recommendations" 
% 
% ^ <beierle@tu-berlin.de> 2017-01-31T14:58:05.897Z.
% * <beierle@tu-berlin.de> 2017-03-21T14:08:21.561Z:
% 
% Reviewer 1 said, we should mention again that the results are from Sowiport. I feel that this i thoroughly explained before and does not need to be mentioned here.
% 
% ^.

\begin{figure}[ht]
	\centering
	\includegraphics[width=1.0\columnwidth]{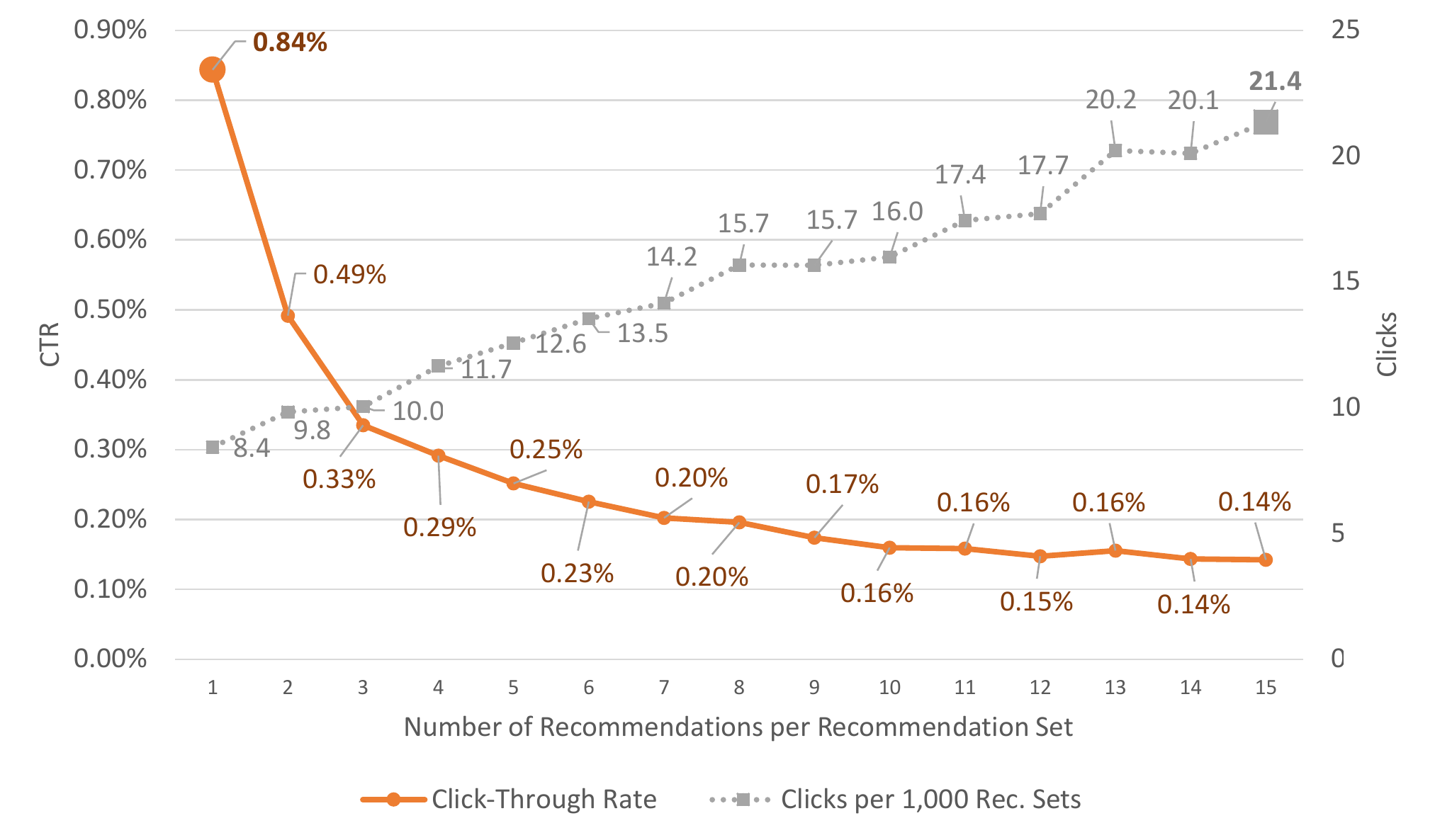}
	\caption{Click-through rates (solid) and average absolute number of clicked recommendations (dashed) with respect to the number of displayed recommendations.}
	\label{fig:ctr}
\end{figure}

The solid orange line in Figure \ref{fig:ctr} shows the CTR by the number of
displayed recommendations.
The higher the number of recommendations, the lower the overall CTR is.
The dashed line shows the average absolute number of clicked
recommendations (per 1,000 recommendations).
The bigger the recommendation set size is, the higher the number of absolute clicks is.

When only one recommendation was displayed, the CTR was 0.84\% on average. This means, when our recommender system delivered 1,000 times one recommendation each, 8.4 recommendations were clicked. For two displayed recommendations, the CTR was only 0.49\% on average. This means, when our recommender system delivered 1,000 times two recommendations each (2,000 in total), 9.8 recommendations were clicked (half clicked on the first one, half on the second one).
Overall, if one or two recommendations are displayed, does not make a big difference
in the absolute number of clicks, it only increases by 17\% (8.4 to 9.8).
%From one to two displayed recommendations, the absolute number of clicked
%related-articles only increases by 17\% (8.4 to 9.8 clicks per 1,000 recommendation sets).
%
When fifteen recommendations were shown, the CTR was at the minimum of 0.14\% on average,
while the absolute number of clicks was at the maximum of 21.4.

Comparing these results with our expectations given in Section \ref{sec:method},
we can see that the CTR unexpectetly decreases rapidly and has a clear
maximum at one displayed recommendation.
Furthermore, contrary to our expectations, the
absolute clicks keep increasing instead of having a maximum value at a few
recommendations.
%ADDED for camera-ready end
The results show an under-proportional increase in average clicks on the displayed
recommendations.
Displaying twice as many recommendations does not double the clicks.
In order for the absolute number of clicked recommendations (per 1,000) to double
from 8.4 (for one displayed recommendation) to 17, the number of displayed
related-articles has to be raised to 10 or 11.
When 15 recommendations are displayed, only 2.5 as many recommendations are clicked compared
to displaying a single recommendation.
Regarding choice overload this implies
that having more recommendations to choose from does, in general,
only create a small incentive for the user to click on more of the
displayed recommended items.
There are some points to consider when interpreting these results.
Many documents in Sowiport only have sparse information,
%, see for example Figure \ref{fig:sparseinfo},
thus they might not be interesting for the users.
Another possible option why the experimental results are so unclear about
the number of recommendations to give is that the
relevance of the recommendations might have been too low, so
that many users did not click further recommendations after clicking the first one.
In that case, the research should be repeated when we are able to deliver better recommendations.
%
%
%
%\begin{figure}[ht!]
%	\centering
%	\includegraphics[width=0.75\columnwidth]{figures/sparseinfo.png}
%	\caption{Screenshot of the Sowiport Digital Library showing very sparse information about a document.}
%	\label{fig:sparseinfo}
%\end{figure}
%
The session-length is another aspect to consider.
For instance, if one user visits two pages and on gets 15 recommendations on each page,
we assume the CTR will be higher than for a user who looks at 10 pages and gets 15
recommendations on each page.
An additional aspect of sessions is that so far, we did not filter for recommendations
that have already been shown to a user.
So, if a  user looks at 15 detail pages and gets 10 related-article recommendations on each,
most likely there will be duplicate recommendations, and hence the CTR decreases the more
recommendations are shown.
Another aspect to consider in the results' interpretation is
how users might use Sowiport.
If the user clicks on a recommendation, a new tab is opened.
If the recommendation was good, she might forget about the other open tab
-- especially if there are further good recommendations shown in the new tab.

%
%Finally, we want to look
%at the average CTRs by used recommendation algorithm that calculated
%the displayed recommendations, see Figure \ref{fig:recalgo}.
%We can see that
%content-based filtering achieves a 43\% higher CTR than
%random recommendations (0.20 \% vs. 0.14\%).
%
%\begin{figure}[H]
	%\centering
	%\includegraphics[width=0.65\columnwidth]{figures/recalgo3.pdf}
	%\caption{Average click-through rate by recommendation algorithm.}
	%\label{fig:recalgo}
%\end{figure}

\section{Conclusion and Future Work}
% * <beierle@tu-berlin.de> 2017-03-21T14:11:40.665Z:
% 
% From Review 1: "What consequences have these results for information retrieval/searching in general?"
% 
% The results could be valid in other fields that use recommender systems as well. I'm not sure if/where this fits in the paper though.
% 
% 
% 
% From Review 1: "What concrete recommendations can the authors give DL operators who provide document-related recommendations?"
% 
% I am not sure how to address this as I think we do not have concrete suggestions yet.
% 
% ^.
The average clicks on displayed recommendations
under-proportionally increase with the number of displayed items.
In order for the clicks to double, the size of the
recommendation set has to be increased from one to 10 or 11.
These numbers might imply that the users quickly feel
overloaded by choice.
The results differ from our expectations.
Just based on these numbers, we could conclude that we should only display
one recommendation because the CTR is highest -- or to keep displaying
even more than 15 recommendations until the number of absolute clicks does
not increase anymore.
Further research will be necessary to determine a good number of recommended
items to display.
%
%Investigating the CTRs by utilized recommendation algorithm,
%we showed that
%content-based filtering yields significantly higher CTRs than random recommendations,
%indicating that the CTR is not only dependent on the number of displayed recommendations
%but also on the relevance of the recommended related-articles.

%- at least for the domain of digital libraries -
%the quality of recommendations seems to have a bigger impact
%than the number of displayed recommendations.

Our results are based on Sowiport.
Further research is necessary to confirm if our findings also apply to other digital libraries.
We therefore plan to repeat our research, for instance, with JabRef and the library of the
Technical University of Munich.
Future work could also include using other evaluation methods and metrics than CTR (e.g., a user study, or user ratings, or tracking which recommended item were actually exported or saved) or making a survey and asking the operators of the other digital libraries how they decided the number of displayed recommendations.
Furthermore, a question to be discussed is what recommender systems in digital libraries should
try to achieve, e.g., maximizing CTR, maximizing the number of clicked recommendations, etc.

\subsubsection*{Acknowledgments.}
This work has received funding from
%the European Union’s Horizon 2020 research and innovation program under grant agreement No 645342, project reTHINK and from
project DYNAMIC\footnote{\url{http://www.dynamic-project.de}} (grant No 01IS12056), which is funded as part of the Software Campus initiative by the German Federal Ministry of Education and Research (BMBF). This work was also supported by a fellowship within the FITweltweit programme of the German Academic Exchange Service (DAAD). This publication also has 
emanated from research conducted with the financial support of Science Foundation Ireland (SFI) under Grant Number 13/RC/2106.
We are further grateful for the support provided by Sophie Siebert.
%TODO: ack checken

%
% ---- Bibliography ----
%
%\begin{thebibliography}{5}
%
%\bibitem {clar:eke}
%Clarke, F., Ekeland, I.:
%Nonlinear oscillations and
%boundary-value problems for Hamiltonian systems.
%Arch. Rat. Mech. Anal. 78, 315--333 (1982)
%
%\bibitem {clar:eke:2}
%Clarke, F., Ekeland, I.:
%Solutions p\'{e}riodiques, du
%p\'{e}riode donn\'{e}e, des \'{e}quations hamiltoniennes.
%Note CRAS Paris 287, 1013--1015 (1978)
%
%\bibitem {mich:tar}
%Michalek, R., Tarantello, G.:
%Subharmonic solutions with prescribed minimal
%period for nonautonomous Hamiltonian systems.
%J. Diff. Eq. 72, 28--55 (1988)
%
%\bibitem {tar}
%Tarantello, G.:
%Subharmonic solutions for Hamiltonian
%systems via a $\bbbz_{p}$ pseudoindex theory.
%Annali di Matematica Pura (to appear)
%
%\bibitem {rab}
%Rabinowitz, P.:
%On subharmonic solutions of a Hamiltonian system.
%Comm. Pure Appl. Math. 33, 609--633 (1980)

%\end{thebibliography}

\bibliographystyle{splncs03}
\bibliography{references}

%\newpage
%\section*{Appendix}
%\input{text/appendix}

\end{document}